\documentclass{aa}

\usepackage[varg]{txfonts}
\usepackage{graphicx}
\usepackage[colorlinks=true,linkcolor=magenta,citecolor=blue,filecolor=cyan,urlcolor=red]{hyperref}

\makeatletter
\renewcommand*\aa@pageof{, page \thepage{} of \pageref*{LastPage}}
\makeatother

\begin{document} 

   \title{Resolving spatial and temporal shock structures using LOFAR observations of type II radio bursts}
        \titlerunning{Resolving spatial and temporal shock structures}

   \author{D.~E.~Morosan \inst{1,2}
           \and
        I.~C.~Jebaraj \inst{1}
        \and
        P. Zhang \inst{3}
        \and
        P. Zucca \inst{4}
        \and
        B. Dabrowski  \inst{5}
        \and
        P. T. Gallagher \inst{6}
        \and 
        A. Krankowski  \inst{5}
        \and
        C. Vocks \inst{7}
        \and
        R. Vainio \inst{1}
        }

   \institute{Department of Physics and Astronomy, University of Turku, 20014, Turku, Finland \\
              \email{diana.morosan@utu.fi}
        \and
             Turku Collegium for Science, Medicine and Technology, University of Turku, 20014, Turku, Finland
        \and 
            Center for Solar-Terrestrial Research, New Jersey Institute of Technology, Newark, New Jersey, USA
        \and
            ASTRON - the Netherlands Institute for Radio Astronomy, Oude Hoogeveensedijk 4, 7991 PD Dwingeloo, the Netherlands
        \and
            Space Radio-Diagnostics Research Centre, University of Warmia and Mazury, R. Prawochenskiego 9, 10-719 Olsztyn, Poland
        \and
            Astronomy \& Astrophysics Section, Dublin Institute for Advanced Studies, DIAS Dunsink Observatory, Dublin D15 XR2R, Ireland
        \and
             Leibniz Institute for Astrophysics Potsdam (AIP), An der Sternwarte 16, 14482 Potsdam, Germany
             }

   \date{Received ; accepted }

 
  \abstract
    {Collisionless shocks are one of the most powerful particle accelerators in the Universe. In the heliosphere, type II solar radio bursts are signatures of electrons accelerated by collisionless shocks launched at the Sun. Spectral observations of these bursts show a variety of fine structures often composing multiple type II lanes. The origin of these lanes and structures is not well understood and has been attributed to the inhomogeneous environment around the propagating shock. }
    {Here, we aim to determine the large-scale local structures near a coronal shock wave using high-resolution radio imaging observations of a complex type II radio burst observed on 3 October 2023. }
    {By using inteferometric imaging from the Low Frequency Array (LOFAR), combined with extreme ultraviolet observations, we investigate the origin of multiple type II lanes at low frequencies (30--80~MHz) relative to the propagating shock wave. }
    {We identify at least three radio sources at metric wavelengths corresponding to a multi-lane type II burst. The type II burst sources propagate outwards with a shock driven by a coronal mass ejection. We find a double radio source that exhibits increasing separation over time, consistent with the expansion rate of the global coronal shock. This suggests that the overall shock expansion is nearly self-similar, with acceleration hotspots forming at various times and splitting at a rate proportional to the shock's expansion. }
    {Our results show the importance of increased spatial resolution in determining either the small-scale spatial properties in coronal shocks or the structuring of the ambient medium. Possible shock corrugations or structuring of the upstream plasma at the scale of 10$^5$~km can act as hotspots for the acceleration of suprathermal electrons. This can be observed as radiation that exhibits double sources with increasing separation at the same expansion rate as the global shock wave.   }


\maketitle


\section{Introduction}

{Collisionless shocks are among the most fundamental phenomena in the Universe, responsible for some of the most visually spectacular cosmic objects. They efficiently dissipate energy across various spatial and temporal scales, enabling the heating and acceleration of particles \citep[][]{Sagdeev66,Galeev76,Kennel85}. The heliosphere provides an ideal laboratory to study naturally occurring collisionless shocks, such as the Earth's bow shock, which has been investigated extensively through observations and numerical simulations \citep[e.g.][]{lembege04,Krasnoselskikh13,johlander2016}. These studies reveal that collisionless shocks are not rigid, planar structures but are highly structured and deformed at scales ranging from sub-ion to large magneto-hydrodynamic scales.}

{Recent in situ observations from Parker Solar Probe \citep[PSP;][]{Fox2016} have shown that the structure of strong shocks near the Sun is highly variable but generally aligns with theoretical predictions \citep[e.g.][]{jebaraj2024}. However, the formation and evolution of shocks in the highly magnetized and dense lower solar corona remain poorly understood due to the lack of in situ measurements. The spatial and temporal variability of shocks or upstream plasma in the low corona is largely unknown, leading to a limited understanding of how shocks, such as those studied by \cite{jebaraj2024}, evolve and accelerate particles in this region. In the absence of in situ observations in the lower corona, remote-sensing observations become essential for gaining critical insights into the complex evolution of shocks.}

   \begin{figure*}[ht]
   \centering
  \begin{minipage}[c]{0.73\textwidth}
          \includegraphics[width=12.9cm]{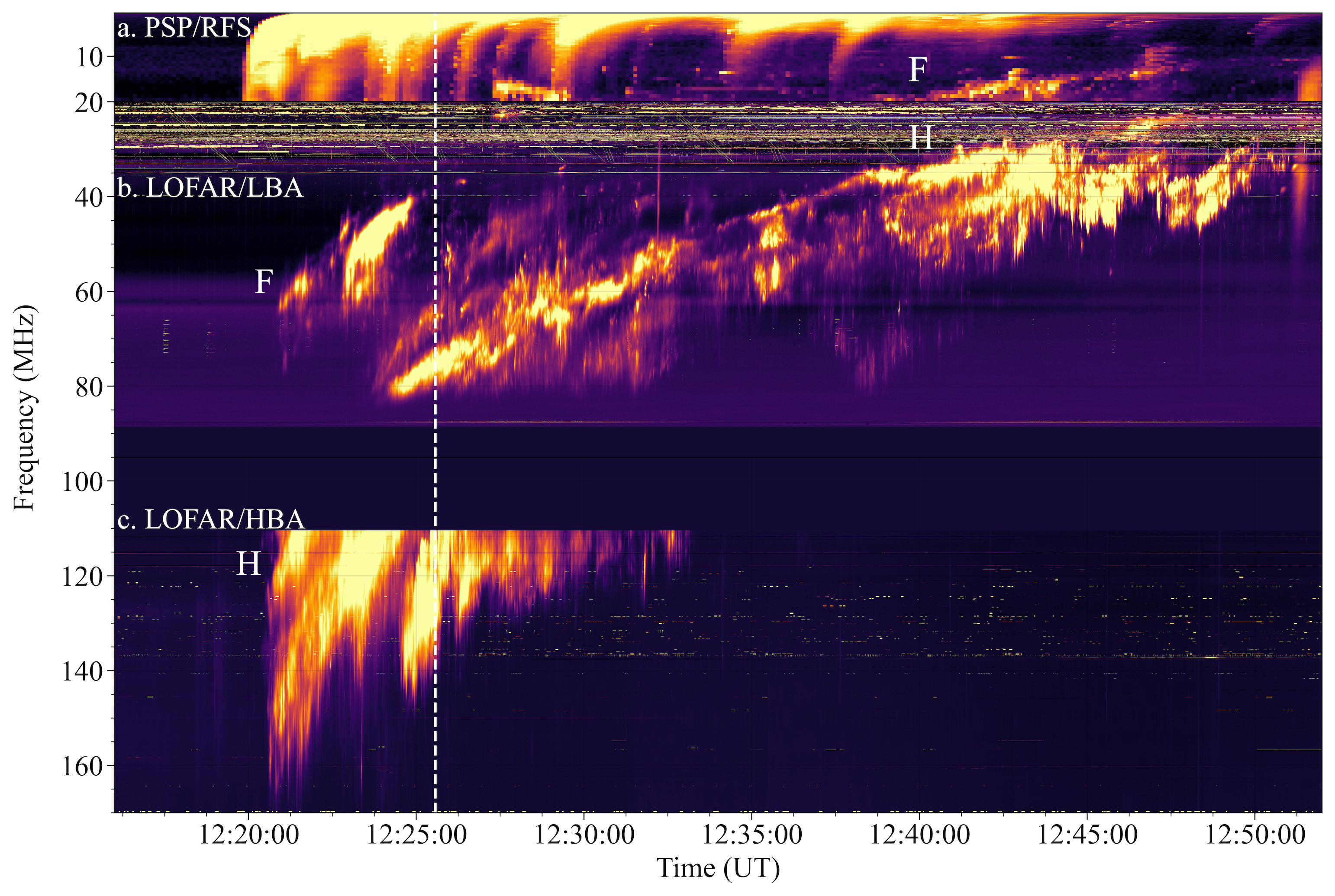}
  \end{minipage}\hfill
   \begin{minipage}[c]{0.27\textwidth}
      \caption{Combined dynamic spectrum of the type II radio burst observed on 3 October 2023. The dynamic spectrum is composed of spectra from PSP/RFS (a), LOFAR/LBA (b), and LOFAR/HBA (c). The PSP time stamp is shifted to Earth time for comparison with the ground-based LOFAR spectra.  }
         \label{fig:fig1}
    \end{minipage}
   \end{figure*}

{Traditionally, remote-sensing observations of shock waves in the low solar corona have been conducted in extreme ultraviolet (EUV), white light (WL), and radio wavelengths \citep[e.g.][]{vourlidas2003, ma2011, smerd1970}. While EUV and WL observations can identify the presence of a shock, the details of shock evolution are best studied through radio wavelengths. Suprathermal electrons accelerated by a propagating coronal shock undergo beam-plasma instabilities, generating type II radio bursts. These emissions depend on the ambient electron density (\(n_\mathrm{e}\)) and are generated at the plasma frequency (\(f_\mathrm{pe} = 8980 \sqrt{n_\mathrm{e}} ~\mathrm{Hz}\) where \(n_\mathrm{e}\) is in cm$^{-3}$) and its harmonics. Due to the radial gradient of \(n_\mathrm{e}\) in the solar corona and the motion of the shock over time, type II bursts exhibit a frequency drift with respect to time. This drift is reflected in frequency-time spectrograms (dynamic spectra hereafter) and can show spatial and temporal changes in both the ambient medium and the propagating shock wave \citep[e.g.][]{zhang2024b}. Since the phenomena that lead to these emissions vary in space and time, in addition to dynamic spectra, radio imaging is essential to obtain spatial information on the emission. Recent advancements in radio imaging instrumentation such as the Low Frequency Array \citep[LOFAR;][]{lofar2013} have made simultaneous sub-second, high-frequency resolution imaging and spectroscopy possible to track the drift evolution of a type II burst. }

{Imaging of type II bursts has recently revealed multiple radio sources at spatially separated regions \citep[e.g.][]{mo19a,zhang2024}. This is attributed to multiple hot spots for suprathermal electron acceleration at the shock \citep[e.g.][]{mo19a, morosan2024}. The most commonly discussed mechanism for a shock to produce an accelerated beam of electrons is through adiabatic reflection in the presence of a steep magnetic gradient, which is possible when the shock wave propagates nearly perpendicular to the ambient magnetic field \citep[][]{Leroy84, mann2018, jebaraj2023}. From a global perspective, a shock driven by a solar transient in the low corona would have such a geometry primarily in its lateral regions \citep[e.g.][]{jebaraj2021,morosan2022a}. However, this geometry could also be provided locally by spatial deformations of the shock front or small-scale structures in the upstream \citep[e.g.][]{zlobec1993, morosan2024}. }

\begin{figure*}[!ht]
\centering
    \includegraphics[width=0.95\linewidth]{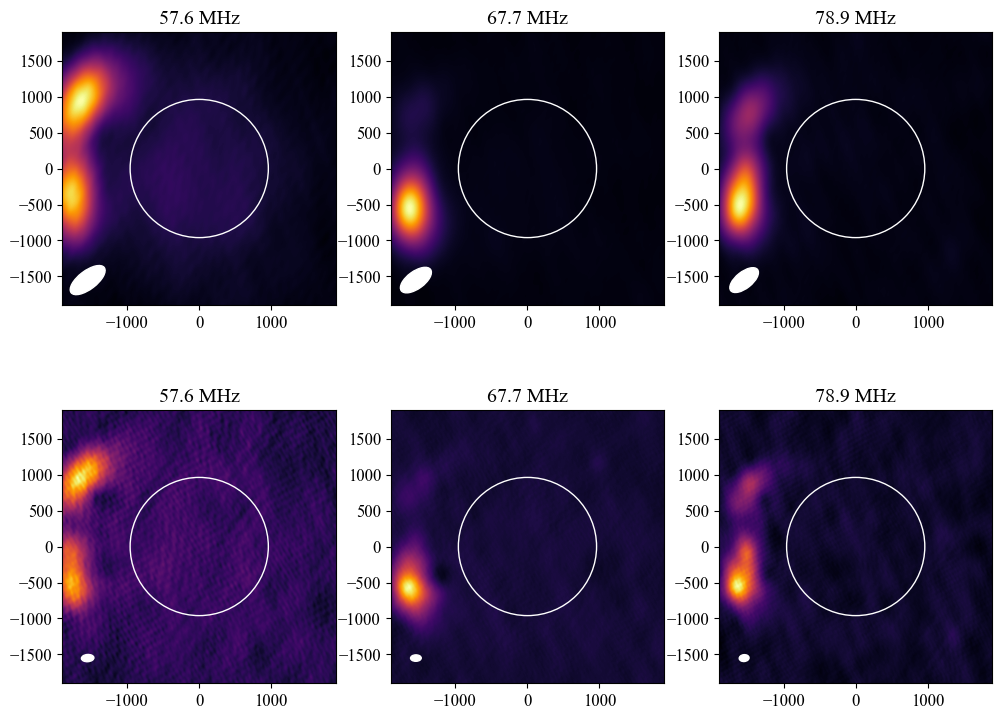}
    \caption{Radio images of the type II burst using two different baselines. The images are at 12:25:34~UT, time which is marked as a white dashed line in the dynamic spectrum in Fig.~\ref{fig:fig1}, and cover three frequency subbands: 57.6, 67.7 and 78.9~MHz, from left to right, respectively. The top panels make use of a baseline of 15 and 10~km, while the bottom panels have a larger baseline of 20 and 32~km, in the East-West and north-south directions, respectively. The size of the beam corresponding to each frequency subband and baseline is shown in the bottom left corner of each panel. In this paper, we present a comprehensive analysis on the spatial properties of a multi-lane type II radio burst and associated shock wave using new and unprecedented high-resolution radio imaging observations with LOFAR. }
    \label{fig:fig2}
\end{figure*}

{Since type II fine structures reveal new aspects of the spatial and temporal evolution at the shock, in this study, we focus on features known as type II multi-lanes. It is a phenomenon where a type II burst consists of two or more lanes, that may have similar drift rates but not necessarily the same spectral morphology \citep[][]{zimovets2015, alissandrakis2021}. Multi-lanes with similar spectral morphology are often known as split-bands \citep[e.g.][]{vr01}. Recent radio imaging studies have demonstrated that multi-lanes in type II bursts originate from different locations at the shock \citep[e.g.][]{bhunia2023, morosan2023}. Given that these lanes can have somewhat similar characteristics, this would imply that they are generated in regions of the shock that are in close proximity to each other. If this is the case, then studying the variations in the temporal and spatial drift of individual type II lanes should provide clues on how two or more emission regions in close proximity to each other evolve. }

\begin{figure*}[ht]
\centering
    \includegraphics[width=0.8\linewidth]{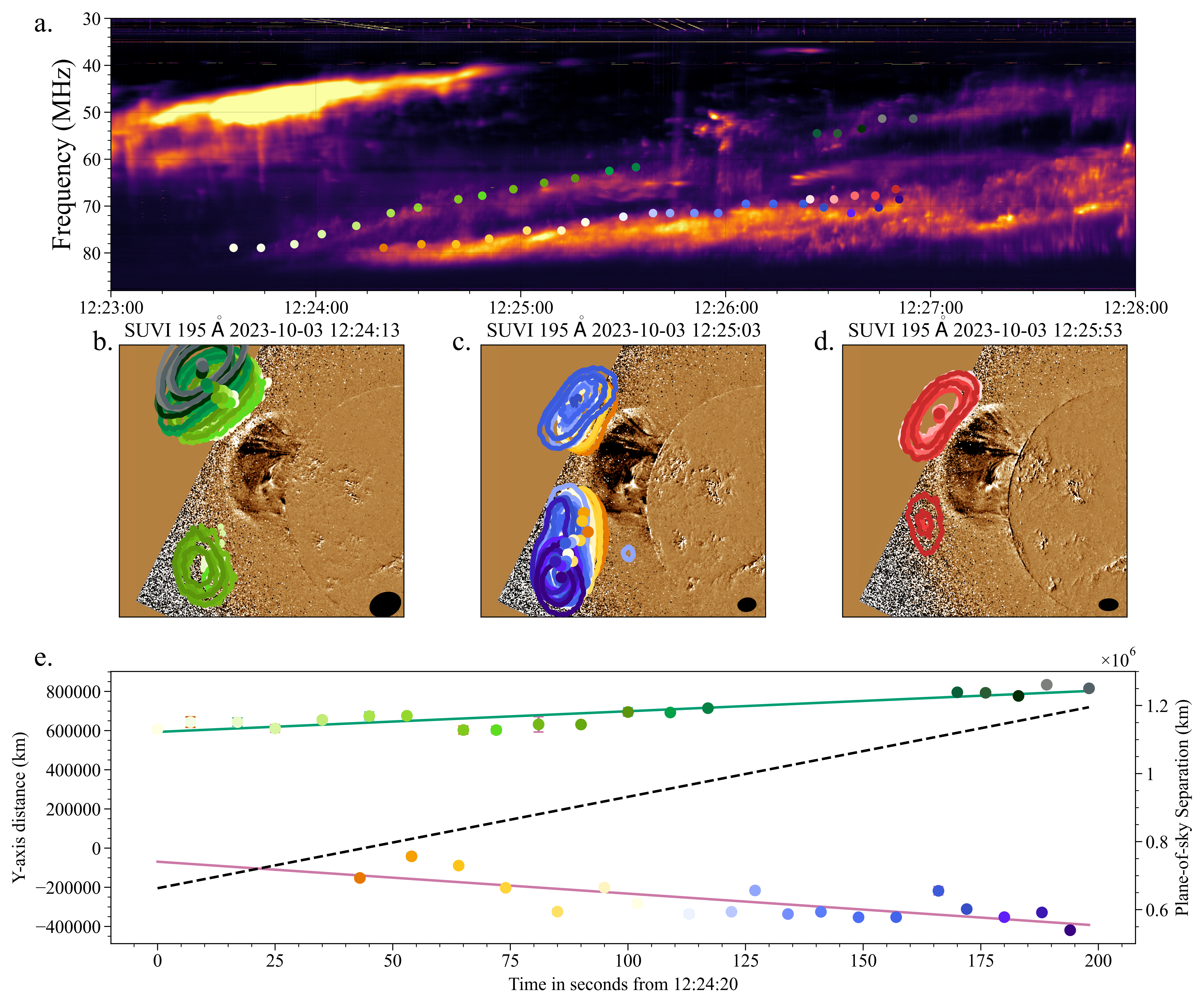}
    \caption{Zoomed-in dynamic spectrum of the type II burst and images of the source location along the leading edge of the two lanes. The top panel shows the start of the harmonic type II lanes in the LBA dynamic spectrum. The color-coded dots track the leading edges of the two lanes. The middle panels show the color-coded contours of the source locations for both lanes overlaid on GOES / SUVI 195~\AA\ running difference images of the CME eruption. The contour levels are at 50 and 70\% of the maximum intensity in each image. The bottom panel shows the Y-axis distance between the two radio sources using the same colour-coding and their separation as a black dashed line.}
    \label{fig:fig3}
\end{figure*}


\section{Observations} \label{sec:analysis}

{On 3 October 2023, a long-duration type II radio burst was recorded with LOFAR as part of the regular cycle observations (LC20\_001) using two simultaneous modes: interferometric mode \citep[e.g.][]{mann2018b} and tied-array beams mode \citep[e.g.][]{morosan2022b}. The type II burst started at 12:21~UT (Earth time) lasting for $\sim$40~minutes (see Fig.~\ref{fig:fig1}). The type II was first observed in LOFAR's High Band Antennas (HBAs) starting at a frequency of 170~MHz and continued to drift to low frequencies to be observed in the Low Band Antennas (LBAs). Type II bursts were also observed by the Radio Frequency Spectrometer \citep[RFS;][]{rfs2017}. In Fig.~\ref{fig:fig1}, the times were shifted to 1 AU for direct comparison between the two different observatories: PSP and LOFAR. PSP was approximately in quadrature to Earth and at a closer distance to the Sun (0.25~AU). PSP also has a better view of the  eruption, which originated on the far-side of the visible solar limb. As a result, additional radio bursts (for example type III bursts that are not seen from Earth) can be observed with PSP. The radio bursts observed (including the type II in the present study) were analysed in relation to the far-side eruption in \citet[][]{morosan2025}. In the present study, we focus on the smaller-scale spatial and spectral characteristics of the multi-lane type II burst. The LBA observations show a complex type II structure with multiple lanes (see Fig.~\ref{fig:fig1}). Much of the emission in the LBA band is likely harmonic since additional type II lanes were observed by PSP between 10-20~MHz with an approximately 2:1 frequency ratio to the LBA type II lanes (see Fig.~\ref{fig:fig1} where fundamental and harmonic lanes are labelled F and H, respectively). }

{LOFAR interferometric images were recorded in the LBA band with 60 frequency subbands available for imaging during the entire observation period from 11:30 to 13:29~UT and with a time cadence of 0.167~s. We used a baseline of up to 32~km in the East-West direction for this study that includes the core and some of the remote LOFAR stations in the Netherlands. The furthest remote station used in the observational setup was excluded due to the significant phase differences to the other stations resulting in the 32~km basline. For comparison, we also generated images using a shorter East-West baseline of 15~km, where the farthest five remote stations have been removed. The interferometric visibilities were calibrated using the LOFAR Initial Calibration Pipeline \citep[e.g,][]{linc2019}. The calibrated visibilities were then processed with \texttt{wsclean} \footnote{wsclean \href{https://gitlab.com/aroffringa/wsclean}{https://gitlab.com/aroffringa/wsclean}} \citep{offringaWsclean2014} to produce science-ready images. Due to bad ionospheric conditions, the type II radio sources and the Quiet Sun were offset from their true position in the interferometric images and were continuously moving over time periods of minutes throughout the observations. As a result, the images were corrected for ionospheric refraction by re-centering the Quiet Sun over all the subbands for select time steps. These time steps were chosen so that the type II emission was either absent or faint and the Quiet Sun outline was visible to have a Gaussian fitted over its extent. The centroids of this Gaussian were used to re-center the image. Only the time steps closest to the Quiet Sun images can be corrected, thus, only select time steps can be chosen to image the type II bursts and this includes the onset time of the multi-lanes that are the focus of this study. The type II lanes were then imaged at several frequencies and their trajectory can be tracked over time. }


\section{Results} \label{sec:results}

   \begin{figure*}[ht]
   \centering
          \includegraphics[width=\linewidth]{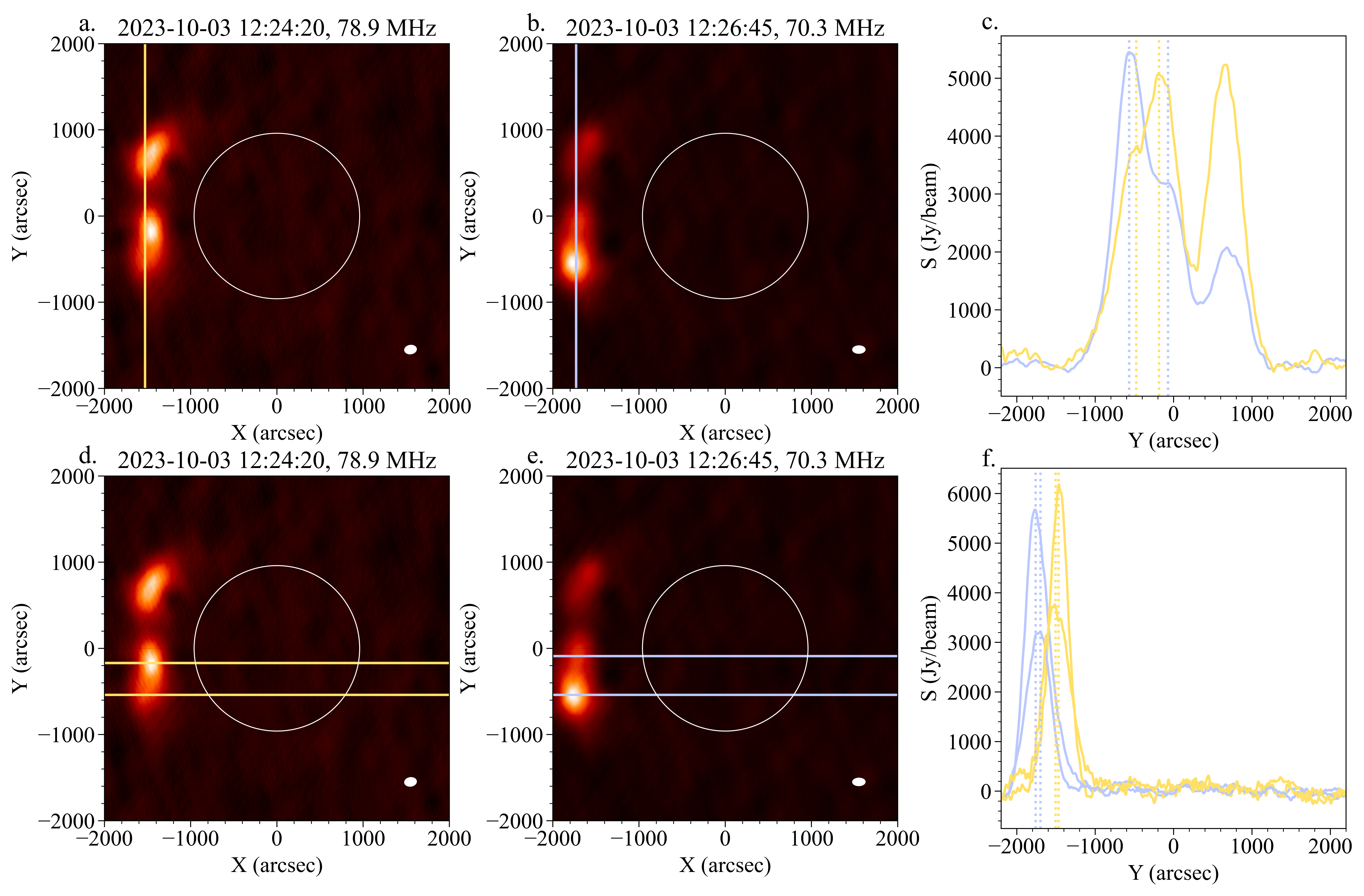}
      \caption{Images and location of the double radio source. (a)-(b), (d)-(e) Interferometric images of the type II radio sources at two different frequencies, 78.9 and 70.3~MHz, and two different times, 12:24:20 and 12:26:45, respectively. (c) Flux of radio sources extracted from the slits marked in panels (a) and (b). (f) Flux of radio sources extracted from the slits marked in panels (d) and (e). The local maxima in the case of the double source are denoted by the vertical dotted lines in panels (c) and (f). 
    }
   \label{fig:fig4}
\end{figure*}


\begin{figure}[ht]
\centering
    \includegraphics[width=0.95\linewidth]{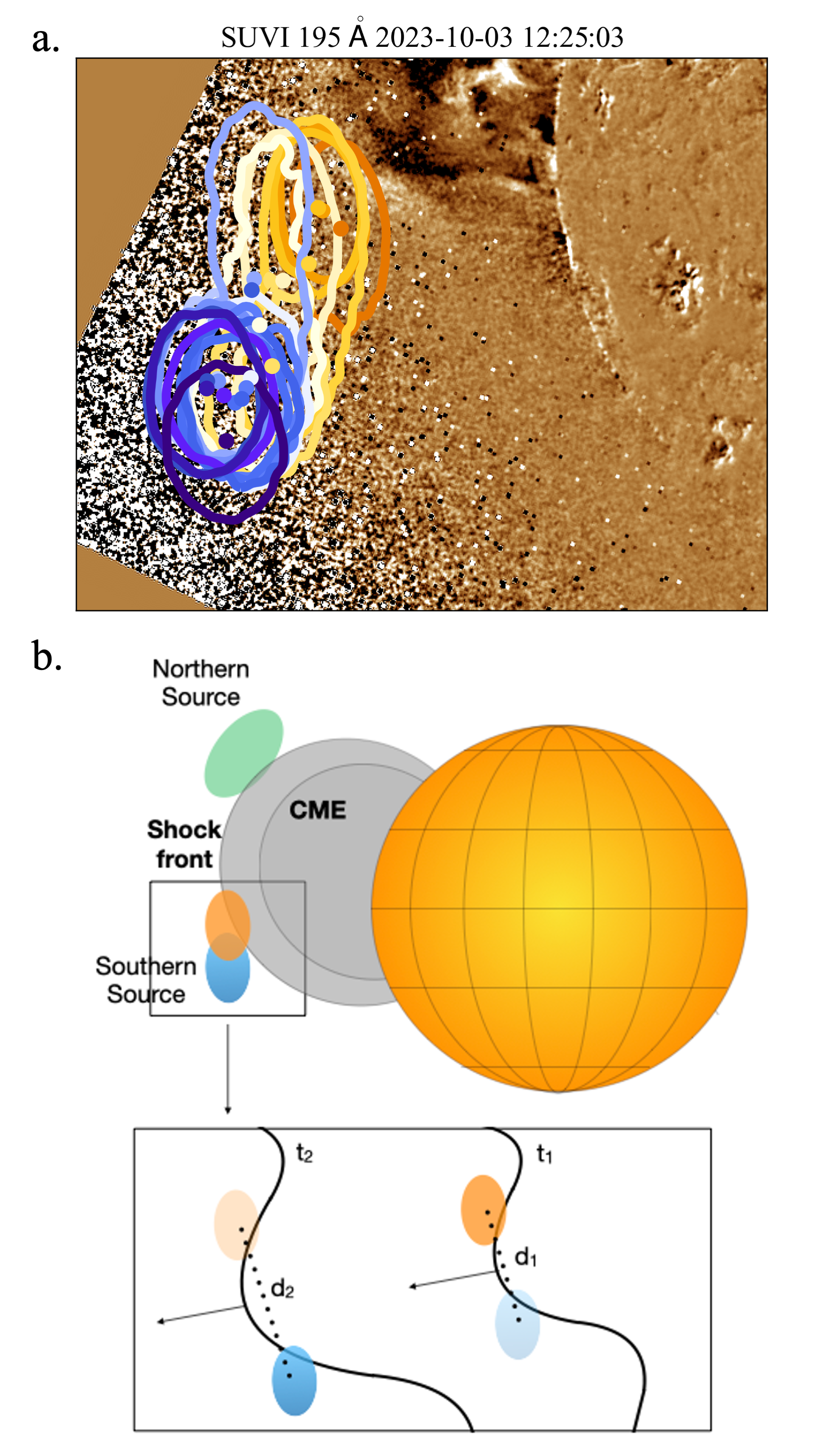}
    \caption{Locations and shock geometry of the double type II sources. (a) Radio contours of the double source at 70\% of the maximum intensity level in each image, overlaid on a GOES / SUVI 195~\AA\ running difference image. (b) Cartoon showing the location of the type II bursts and local shock geometry that is likely to explain the double source separation. As the non-uniform shock front expands between times $t_1$ and $t_2$, the separation between the radio sources increases from $d_1$ to $d_2$.}
    \label{fig:fig5}
\end{figure}

\subsection{Location and characteristics of the type II lanes }

{The onset of the type II and the multi-lanes composing it can be imaged in the frequency range of 90--30~MHz. An example of radio images over three sub-bands is shown in Fig.~\ref{fig:fig2}, using two separate baselines:  a shorter baseline of 15~km in the East-West direction and 10~km in the north-south direction (top panels) and the largest baseline possible in this observation of 20~km in the East-West direction and 32~km in the north-south direction (bottom panels). In the short baseline images, two radio sources are visible located approximately North and South of the central meridian outside the visible solar limb in the plane-of-sky. The radio sources are roughly approximated by the shape of an elliptical Gaussian. The increase in baseline in the bottom panels of Fig.~\ref{fig:fig2} and thus, the increase in the spatial resolution of the images, has a significant effect on the imaging results as the radio sources show additional structures and shapes that start to deviate from an elliptical Gaussian. These new structures and shapes are likely to be overlooked when using shorter baselines, as evident in the top panels of Fig.~\ref{fig:fig2}. For example, the southern radio source that appears as an elongated single source in the low-resolution images, is actually composed of two separate but neighbouring radio sources and this is evident at three different frequencies: 57.6, 67.7 and, 78.9~MHz. In total, there are three radio sources clearly resolved in this time period. }

{The leading edge of the type II lanes is traced in a zoomed-in dynamic spectrum in Fig.~\ref{fig:fig3}, where the colour-coding changes with time and frequency. Here, we focused on the two main type II harmonic lanes. However, we note that a multitude of other emission lanes are present in the spectrum. The color-coded contours of the source locations in the dynamic spectra are shown Figs.~\ref{fig:fig3}b--d for the lower frequency lane and the higher frequency lane, respectively. These contours are overlaid on 195-\AA\ running difference images (separated by one minute) from the Solar Ultraviolet Imager \cite[SUVI;][]{suvi2022} aboard Geostationary Operational Environmental Satellites (GOES) of the coronal mass ejection (CME) driving the shock at three consecutive times. The centroids of the radio emission are also overlaid on this image using the same colouring.  }

{We can make a number of comments about the source locations in the middle panels of Fig.~\ref{fig:fig3}. Firstly, the emission corresponding to the two lanes are generated at different regions of the shock. They can be categorized as northern and southern radio sources, each one corresponding to a different lane, that propagate in north-easterly and south-easterly directions, respectively, away from each other. By extracting the centroids of the radio sources over time, we estimate that the northern and southern sources propagate outwards at similar speeds of 1420 and 1490~km/s, respectively.  In addition to the outward motion, we also perform a linear regression fit for both the x and y-axis separations shown in Fig.~\ref{fig:fig4} (note that only the y-axis separation is shown in Fig.~\ref{fig:fig3}e as it's the most prominent). These fits now allow us to estimate the rate at which they separate from each other with respect to time shown as the dashed line in Fig.~\ref{fig:fig3}e, which is $\sim$2730~km/s. }

{Secondly, within the radio contours of the southern source, we identify two distinct smaller sources, corresponding to the brighter, higher-frequency type II lane in Fig. \ref{fig:fig3}a. The leading edge of this higher-frequency lane also shows a non-uniform drift in the dynamic spectrum. The radio source contours, as shown in Figs. \ref{fig:fig3}c and d, reveal that part of this lane actually originates from the northern source, as indicated by the split in location between the red and blue contours around 12:26:30 UT. The red imaging contours correspond to the red circles in the dynamic spectrum  Fig.~\ref{fig:fig3}d, representing the northern source, which shows the brightest emission at that time and frequency, while the blue contours and circles represent the southern source  Fig.~\ref{fig:fig3}c. Although some red contours are present in the south, they are 30\% fainter. Thus, while the higher-frequency type II lane may appear as a distinct emission in the dynamic spectrum, imaging reveals that multiple sources contribute to its overall morphology. Without radio imaging to identify these multiple sources, tracing the leading edge on the radio spectrum becomes challenging and non-trivial. }

{Thirdly, the neighbouring southern sources are initially indistinguishable and may be mistaken for an extended radio source. However, their spatial separation increases with time and eventually two clearly distinguishable radio bursts are visible. In order to quantify the rate of separation of the two distinct southern radio sources, we extract the radio flux inside a slit applied to the interferometric images (see Fig.~\ref{fig:fig4}). Figures~\ref{fig:fig4}a and b show two interferometric images corresponding to the first contour (orange) and the last contour (blue) of the type II points in the dynamic spectrum in Fig.~\ref{fig:fig3}. The slit is denoted by the vertical lines passing through the radio source in each of these panels. Fig.~\ref{fig:fig4}c shows the peaks in burst intensity along the slit. The first thing to notice is the clearly separated northern and southern sources, that appear as two distinct peaks. However, the first peak corresponding to the southern source shows two local maxima separated by a distance of 0.3~R$_\odot$ at 12:24:20~UT (orange profile) and 0.5~R$_\odot$ at 12:25:45~UT (blue profile). In figs.~\ref{fig:fig4}d and e the slit is along the x-direction and it is denoted by the horizontal lines passing through the radio source in each of these panels. The offset in the x-direction of the two sources is small and can be seen in Fig.~\ref{fig:fig4}f as the small distances between the blue and yellow dashed lines, respectively. The split between the two radio sources is also evident when plotting only the 70\% contour levels of the bursts that is shown in Fig.~\ref{fig:fig5}a. Initially, the higher burst is brighter (orange contours) but as the shock expands the lower source becomes brighter (blue contours), with their separations also increasing over time as outlined in the cartoon in Fig.~\ref{fig:fig5}b. At frequencies around 75~MHz both sources have similar flux levels and appear as one elongated radio burst.  }

\subsection{Expansion speed of the type II radio sources }

{The separations between the northern and southern sources and also the two neighbouring radio sources in the south increase over time. To quantify this change over time, we assume that the projected radio sources obtained via interferometry lie on a circle in the plane of the sky, with the radius of this circle measured from the center of the CME eruption (see Fig.~\ref{fig:fig6}). The simplest way to solve for the rate of change in this radius is by using the relationship between the source distance, $D$ and the angle, $\theta$:

\begin{equation}\label{eq:1}
\frac{D}{2} = r \cdot \sin (\theta/2),
\end{equation}

\noindent where the separation, $D$, is related to the emission centroid x- and y-coordinates by the following relation:

\begin{equation}\label{eq:1.1}
D = \sqrt{ \Delta x^2 + \Delta y^2}.
\end{equation}

In the case of the northern and southern radio sources, we used emission centroids to estimate the coordinates. However, for the double source, the centroids were not very well defined due to their proximity so we used the local maximum inside a slit going through the sources in the x- and y-directions (Fig.~\ref{fig:fig3}). We note that the y-separation is the most prominent. 

\begin{figure}[!ht]
\centering
    \includegraphics[width=\linewidth]{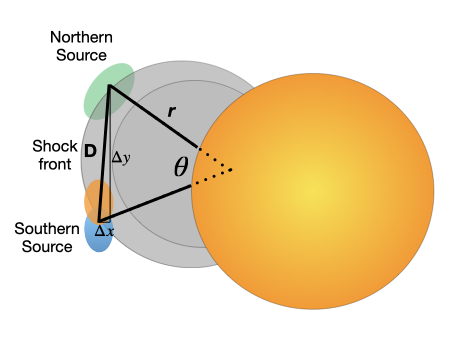}
    \caption{Cartoon highlighting the relation between source separation and angle from the origin of the flare. The source separation, $D$, is used to calculate the expansion rate by using the angle between the two radio sources, $\theta$, with its origin as the flaring site, and the radial distance, $r$, from the radio source to the flaring region (see Equation~\ref{eq:1}).}
    \label{fig:fig6}
\end{figure}

From data we have obtained that the northern and southern sources are separated by \(\approx 70^{\circ}\) and have a separation rate of 2730~km/s determined from Fig.~\ref{fig:fig3}e. If a shock were to expand self-similarly, then this separation rate is related to the global expansion speed of the shock \citep[e.g.][]{normo2024}. Thus using Equation~\ref{eq:1}, the rate of separation can be related to shock expansion speed as,

\begin{equation}\label{eq:2}
    \frac{dr}{dt} = \frac{1}{2 \sin (\theta / 2)} \frac{dD}{dt}.
\end{equation}

{Substituting the values for \( \frac{dD}{dt} = 2730 \) km/s, and \( \theta =  70^{\circ} \) in Equation~\ref{eq:2}, we can solve for the global rate of change in the radius \( \frac{dr}{dt} \approx 2400 \)~km/s. }

{Similarly, we can also estimate the local rate of change in shock expansion speed from the split in the southern source. The split has an angle \(\theta = 20^{\circ}\) and a separation rate of \( \frac{dD}{dt} \approx 1050 \) km/s. Using this in Eq. \ref{eq:2}, we obtain the local expansion rate, \( \frac{dr}{dt} \approx 3000 \)~km/s. We note that these separation speeds are larger than the plane-of-sky CME speed and this may be due to a number of reasons outlined in Appendix~\ref{app:a}.}


\section{Discussion and conclusion} \label{sec:discussion}

{In this paper, we demonstrated that increasing spatial resolution in interferometric radio observations significantly impacts the observed radio source size and structure. With longer baselines, the radio sources appear smaller and reveal more complex features, such as a double source. Without high-resolution imaging, this double source could easily be misinterpreted as a single radio burst. These findings highlight that relying solely on dynamic spectra to interpret complex bursts such as type II bursts can lead to misconceptions about the underlying physics of the source.  }

{The most important finding in this study is the double radio source that exhibits increasing separation over time, consistent with the global separation rate of the northern and southern radio sources. This suggests that the overall shock expansion is nearly self-similar, with acceleration hotspots forming at various times and splitting at a rate proportional to the shock's expansion. The end result of these acceleration hotspots could be observed as radiation that exhibits double or multiple sources with increasing separations when imaged, propagating at a similar rate as the global shock wave. A significant uncertainty regarding shocks in the corona is indeed whether they evolve in a self-similar manner, as the lack of 3D observations makes this difficult to confirm. It is often thought that irregularities or corrugations in the medium can deform the shock surface, potentially causing deviations from self-similar behavior in 3D \citep[see for example the 3D shock simulation in][]{wijsen2023}. The nature of the double source suggests that either the radio emission or electron acceleration is inhibited in the separation region. Changes in acceleration along the shock front can be explained by changes in the local shock geometry, where, at small scales along the shock front, the shock occasionally deviates from a quasi-perpendicular geometry necessary for the rapid and efficient acceleration of electron beams. Below, we discuss two scenarios in which the properties of the shock wave or ambient medium can explain the presence of this double source:}

{The double radio source presented here can be emitted on ether side of a shock corrugation (see Fig.~\ref{fig:fig5}b). Previous studies suggested that a wavy or rippled shock wave is most likely to explain the generation of fine structures in type II emission \citep[e.g.][]{zlobec1993, mo19a}. The cartoon in Fig.~\ref{fig:fig5}b shows the possible shock geometry that can explain the presence of a double radio source, as well as its increasing separation as the shock wave expands. In this case, the separation of the radio sources serves as a proxy for the spatial scale of the corrugation itself. }Our estimations suggest that such a corrugation has a rather large plane of sky size, on the order of $\sim10^5$~km, but much smaller than the shock itself. We note that this scale size may be limited by the LOFAR beam size used for imaging. These deformations are caused by highly variable gradients in density and magnetic field within the ambient medium through which the shock propagates \citep[][]{wijsen2023}. These surface deformations are significantly larger than the tangential surface waves (ripples) generated by super-critical shocks \citep[e.g.][]{Gedalin24}. However, they can manifest at various spatial and temporal scales, depending on the corresponding variations in the ambient medium. }  

{Extreme inhomogeneities in the ambient medium are likely to deform the shock. However, these variations themselves can also be the cause of the double radio source. An example would be the presence of non-radial field lines such that the shock geometry switches from quasi-perpendicular to quasi-parallel and then back again to explain the region of no emission between the two radio sources. These non-radial field lines have been suggested before in the study by \citet[][]{morosan2024} to explain the likelihood of a global oblique shock becoming perpendicular at local scales. The source separation would then give the scale of field line structuring in the low corona. Another possibility, is the presence of large intermittent density structures in the corona \citep[][]{deforest2018}. }

{To conclude, given this variability in shock structure or ambient medium, electron acceleration may occur intermittently, resulting in radio emission with numerous fine structures with varying drifts and brightness within the larger-scale emission from a region of the global shock. These are commonly observed characteristics of type II radio bursts that in turn reflect the complex structural variations of the shock. We have shown that these variations can lead to hotspots for the acceleration of electrons and also expand as the shock propagates outward at a similar rate as the global shock wave. }

%




\begin{acknowledgements}{D.E.M. acknowledges the Research Council of Finland project `SolShocks' (grant number 354409). I.C.J was funded by the Research Council of Finland project SHOCKSEE (grant No.~346902). This study has received funding from the European Union’s Horizon Europe research and innovation programme under grant agreement No.\ 101134999 (SOLER). The paper reflects only the authors' view and the European Commission is not responsible for any use that may be made of the information it contains. The research is performed under the umbrella of the Finnish Centre of Excellence in Research of Sustainable Space (FORESAIL) funded by the Research Council of Finland (grant no. 352847). The authors wish to acknowledge CSC – IT Center for Science, Finland, for computational resources. This paper is based on data obtained with the LOFAR telescope (LOFAR-ERIC) under project code LC20\_001. LOFAR (van Haarlem et al. 2013) is the Low Frequency Array designed and constructed by ASTRON. It has observing, data processing, and data storage facilities in several countries, that are owned by various parties (each with their own funding sources), and that are collectively operated by the LOFAR European Research Infrastructure Consortium (LOFAR-ERIC) under a joint scientific policy. The LOFAR-ERIC resources have benefited from the following recent major funding sources: CNRS-INSU, Observatoire de Paris and Université d'Orléans, France; BMBF, MIWF-NRW, MPG, Germany; Science Foundation Ireland (SFI), Department of Business, Enterprise and Innovation (DBEI), Ireland; NWO, The Netherlands; The Science and Technology Facilities Council, UK; Ministry of Science and Higher Education, Poland. } 
\end{acknowledgements}

\bibliographystyle{aa} 
\bibliography{aanda_main.bib} 

\begin{appendix} 

\section{Radio source separation rate and shock speed} \label{app:a}

\begin{figure}[!ht]
\centering
    \includegraphics[width=\linewidth]{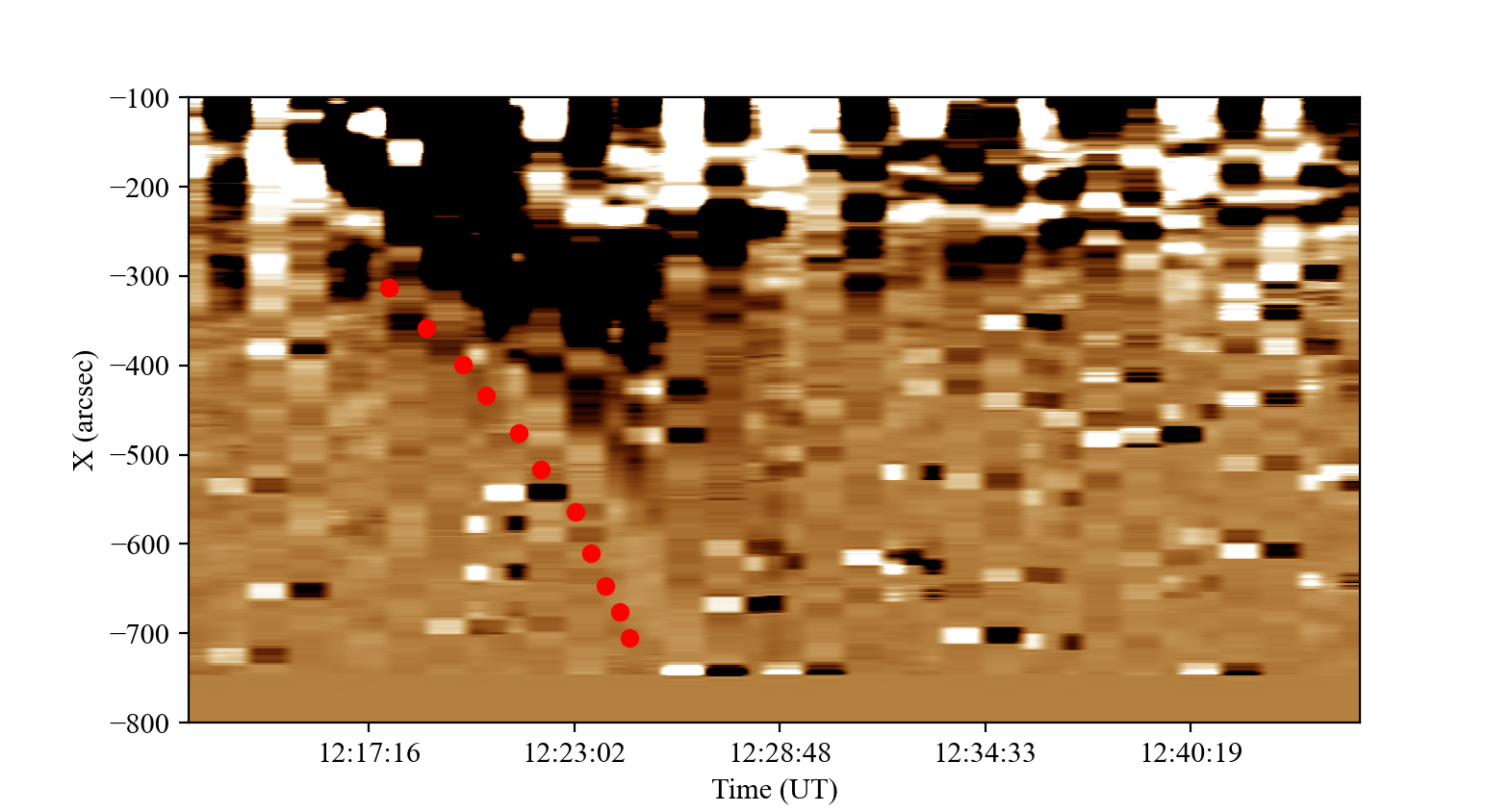}
    \caption{Slit applied to SUVI 195~\AA\ running difference images in the direction of the mid-point of the CME propagating radially outwatds. The red dots denote the faint outer edge of the CME eruption in the EUV images that were used to estimate its radial outwards speed using a linear regression fit.  }
    \label{fig:figA1}
\end{figure}

{The separation rate of the radio sources is related to the global expansion of the shock wave. However, this separation speed may not be exactly the same as the expansion of the shock for a number of reasons. Firstly, the plane-of-sky speed of the CME obtained from SUVI images is only $\sim$680~km/s (see Fig.~\ref{fig:figA1}), and assuming self-similar expansion and the shock corresponding to the outer faint EUV emission in the SUVI images, this speed is also indicative of the expansion rate of the shock. This speed is, however, expected to be a lower limit since the eruption originates behind the visible solar limb, facing away from observers at Earth. Given that the flare originates at a Carrington longitude of 213$^{\circ}$ (see \citealt[][]{morosan2025} where the longitude was estimated from far side observations), thus the angle between the eruption site and the limb (Carrington longitude 177$^{\circ}$) is 36$^{\circ}$, a more accurate radial speed estimate of the shock is:}
\[
v_s =  \frac{680}{ \cos{36^{\circ}} } \approx 840~{\rm km/s}
\]
{However, the above value for de-projected speed makes the assumption that the CME propagates radially outwards from the flaring region which is not always the case. Thus, further projection effects are likely the cause for the discrepancy between the CME speed obtained from EUV images and the shock expansion inferred from radio imaging. Another factor contributing to this discrepancy is that the radio bursts may not necessarily travel along the shock normal but at a certain angle. Then there is a dependence on the actual shock speed and this angle. However, the self-similarity of the expansion remains unaffected when correcting for this angle. These projection effects are, however, not possible to estimate without multi-viewpoint images at both EUV and radio wavelengths.  }

\end{appendix}

\end{document}